\documentclass[aip,reprint,apl]{revtex4-1}
\usepackage{graphicx}
\usepackage{amsmath}
\usepackage{amssymb}
\usepackage{booktabs}
\usepackage{array}
\usepackage{xcolor}
\usepackage{acronym}
\usepackage{pifont}
\usepackage{commath}

\usepackage{type1cm}
\usepackage{eso-pic}
\usepackage{color}
\usepackage{soul}
\usepackage{txfonts}

\begin{document}

%\preprint{}

\title{Electron effective mass in unintentionally doped In$_{0.33}$Ga$_{0.67}$N determined by mid-infrared optical Hall effect} %Title of paper



\author{Nerijus Armakavicius}
\affiliation{Terahertz Materials Analysis Center, Department of Physics, Chemistry and Biology IFM, Link\"{o}ping University, Sweden}
\author{Vallery Stanishev}
\affiliation{Terahertz Materials Analysis Center, Department of Physics, Chemistry and Biology IFM, Link\"{o}ping University, Sweden}
\author{Sean Knight}
\affiliation{Department of Electrical and Computer Engineering and Center for Nanohybrid Functional Materials, University of Nebraska-Lincoln, USA}
\author{Philipp K\"{u}hne}
\affiliation{Terahertz Materials Analysis Center, Department of Physics, Chemistry and Biology IFM, Link\"{o}ping University, Sweden}
\author{Mathias Schubert}
\affiliation{Terahertz Materials Analysis Center, Department of Physics, Chemistry and Biology IFM, Link\"{o}ping University, Sweden}
\affiliation{Department of Electrical and Computer Engineering and Center for Nanohybrid Functional Materials, University of Nebraska-Lincoln, USA}
\affiliation{Leibniz Institute for Polymer Research, Dresden, Germany}
\author{Vanya Darakchieva}
\affiliation{Terahertz Materials Analysis Center, Department of Physics, Chemistry and Biology IFM, Link\"{o}ping University, Sweden}

%\email[]{}
%\thanks{}


\date{\today}

\begin{abstract}
Mid-infrared optical Hall effect measurements are used to determine the free charge carrier parameters of an unintentionally doped wurtzite-structure $c$-plane oriented In$_{0.33}$Ga$_{0.67}$N epitaxial layer. Room temperature  electron effective mass parameters of $m^{*}_{\bot}=(0.205 \pm 0.013)~m_0$ and $m^{*}_{\parallel}=(0.204 \pm 0.016)~m_0$ for polarization perpendicular and parallel to the $c$-axis, respectively, were determined. The free electron concentration was obtained as $(1.7 \pm 0.2)\times 10^{19}$~cm$^{-3}$. Within our uncertainty limits
we detect no anisotropy for the electron effective mass parameter and we estimate the upper limit of the possible effective mass anisotropy is 7$\%$. We discuss the influence of band nonparabolicity on the electron effective mass parameter as a function of In content. The effective mass parameter is consistent with a linear interpolation scheme between the conduction band mass parameters in GaN and InN when the strong nonparabolicity in InN is included. The In$_{0.33}$Ga$_{0.67}$N electron mobility parameters were found to be anisotropic supporting previous experimental findings for wurtzite-structure GaN, InN, and Al$_{x}$Ga$_{1-x}$N epitaxial layers with $c$-plane growth orientation.
\end{abstract}


\maketitle %\maketitle must follow title, authors, abstract and \pacs

\acrodef{OHE}[OHE]{optical Hall effect}
\acrodef{MIR}[MIR]{mid-infrared}
\acrodef{NIR}[NIR]{near-infrared}
\acrodef{UV}[UV]{ultraviolet}
\acrodef{SE}[SE]{spectroscopic ellipsometry}
\acrodef{IR}[IR]{infrared}
\acrodef{TO}[TO]{transverse optical phonon}
\acrodef{LO}[LO]{longitudinal optical phonon}
\acrodef{LPP}[LPP$^{+}$]{coupled longitudinal optical-phonon plasmon}

%\section{Introduction}
%1) III-V nitrides and InGaN\\
%2) Importance of the effective mass and other FCC parameters\\
%3) Effective mass in GaN and InN and factors effecting it.\\
%4) Theoretical and experimental determination of the effective mass in InGaN\\

Variation of indium content in wurtzite-structure In$_{x}$Ga$_{1-x}$N permits tuning its bandgap from the \ac{NIR} (NIR; InN) to the \ac{UV} (UV; GaN) spectral range.\cite{Junqiao} Device heterostructures employing low-In content In$_{x}$Ga$_{1-x}$N alloys ($x<0.3$) are widely used in blue light emitting diodes, laser diodes and vertical-cavity surface-emitting lasers. On the other hand, higher In content ($x>0.3$) In$_{x}$Ga$_{1-x}$N optoelectronic devices  for operation at long wavelengths are still emerging.  The high electron saturation drift velocity of the two dimensional electron gas in In$_{x}$Ga$_{1-x}$N ($\sim$2$\times$10$^7$cm~s$^{-1}$) makes the material also promising for high frequency and high power electronic switch and transistor applications.\cite{ardaravicius_ssat15} Knowledge of free charge carrier properties of In$_{x}$Ga$_{1-x}$N alloys, and their dependencies on carrier density and In content is prerequisite for device design and operation, and for developing new device architectures. However, reports on electron effective mass, concentration and mobility parameters in In$_{x}$Ga$_{1-x}$N are scarce.

Density functional tight binding calculations predict nearly linear variation  of the electron effective mass parameter in wurtzite-structure In$_{x}$Ga$_{1-x}$N as a function of In content with a small bowing parameter of 2.27$\times$10$^{-3}$.\cite{Elfitur} Cyclotron resonance and Shubnikov-de Haas measurements, which can typically be used to experimentally determine the effective mass, require high mobility parameters and thus low temperatures, and/or very high magnetic fields.\cite{millot_prb11} Alloy disorder and relatively high density of defects in alloyed thin films typically reduce the mobility parameters compared to the binary compounds rendering Cyclotron resonance and Shubnikov-de Haas measurements inapplicable to currently available In$_{x}$Ga$_{1-x}$N materials. The determination of free charge carrier effective mass at room temperature or higher (under device operation condition) temperature is still a challenging task. There are only few reports which estimate experimentally the effective mass parameters in In$_{x}$Ga$_{1-x}$N.\cite{Yadav,Eljarrat} Eljarrat~\textit{et al.} estimated an isotropic average of the electron effective mass parameter between 0.14~$m_0$ and 0.16~$m_0$ ($m_0$ is the free electron mass) for multilayer quantum well structures of In$_{0.05}$Ga$_{0.95}$N and In$_{0.2}$Ga$_{0.8}$N layers using a numerical Kramers-Kronig extension of electron energy loss spectra.  Yadav~\textit{et al.} investigated polycrystalline In$_x$Ga$_{1-x}$N layers and estimated isotropically averaged electron effective mass parameters for In contents from 0.4 to 1 combining parameters obtained from electrical Hall effect and optical reflectivity measurements.\cite{Yadav} The reported values vary between 0.13~m$_0$ and 0.42~m$_0$, and carry large uncertainty limits of more than 50$\%$. Furthermore, values determined from the measured plasma frequencies and the estimated energies of Burstein-Moss shifts assuming parabolic bands differ substantially among the reported set.\cite{Yadav} 

It is well known that the conduction band of InN exhibits nonparabolicity, while in GaN the conduction band is nearly parabolic.\cite{Walukiewicz1} Therefore, the electron effective mass increases in InN with increasing free electron density, which was experimentally verified by Hofmann~\textit{et al.} using optical Hall effect measurements.\cite{hofmann_jem08} Due to the uniaxial symmetry of the wurtzite lattice structure, the  electron effective mass parameter at the conduction band minimum in InN is anisotropic with $m_{\perp}^{*}=0.050~m_0$, $m_{\parallel}^{*}=0.037~m_0$.\cite{hofmann_jem08} With increasing carrier concentration the anisotropy vanishes.\cite{hofmann_jem08} For GaN the effective mass parameter at the conduction band minimum exhibits negligible anisotropy below 1~\% with $m_{\perp}^{*}=0.0237\pm0.006~m_0$, $m_{\parallel}^{*}=0.228\pm0.008~m_0$.\cite{Kasic_prb00} With increasing carrier concentration the small anisotropy remains.\cite{Feneberg_1} Feneberg \textit{et al.} used infrared spectroscopic ellipsometry and showed that the electron effective mass in GaN remains constant for free electron concentrations up to to the low $10^{20}$~cm$^{-3}$ range, as expected for a parabolic conduction band of GaN.\cite{Feneberg_1} Hence, conduction band nonparabolicity is anticipated in In$_x$Ga$_{1-x}$N alloys, and which may vary as a function of the In content. To this end, no accurate effective mass and anisotropy determination has been reported yet for any In content $0\,<\,x\,<\,1$.

Measurements of the \ac{OHE} provide access to the free charge carrier parameters, including their anisotropy, in contactless manner and does not require prior knowledge of band curvature and band gap parameters.\cite{Schubert:03,hofmann_jem08,Schubert_OHE} The \ac{OHE} is a physical phenomenon  that occurs in conductive samples upon interaction with electromagnetic waves with mid-infrared to terahertz frequencies and in the presence of an external magnetic field. The OHE is manifested in optical birefringence caused by the motion of the free charge carriers under the influence of the Lorentz force, and  can be measured by generalized ellipsometry. The free charge carrier properties are then obtained from the ellipsometry data analysis based on parametrized physical model.\cite{Schubert_OHE} 

In this work we determine the electron effective mass, mobility and concentration parameters of wurtzite In$_{0.33}$Ga$_{0.67}$N using \ac{MIR} \ac{OHE} measurements.  We discuss the results on the electron effective mass parameter in terms of In content and free electron concentration dependence, and estimate the upper limit of the electron effective mass anisotropy that may remain hidden within our present experimental error bars.

%\section{Experiment}

An In$_{0.33}$Ga$_{0.67}$N layer with a nominal thickness of 300~nm was grown on an AlN template layer on $c$-plane sapphire ($\alpha$-Al$_2$O$_3$) substrate by molecular beam epitaxy. Both AlN and In$_{0.33}$Ga$_{0.67}$N epilayers have wurtzite crystal structure with their $c$-axes perpendicular to the sample surface. The In content was determined using X-ray diffraction measurements. \ac{MIR}-\ac{SE} measurements (IR-VASE, J.A.~Woollam~Co., Inc.) at zero magnetic field were carried out in the spectral range from 300-1200~cm$^{-1}$ with a resolution of 1~cm$^{-1}$, and at 55$^{\circ}$ and 65$^{\circ}$ angles of incidence. \ac{MIR}-\ac{OHE} measurements were performed within the spectral range of 800-1100~cm$^{-1}$ with a resolution of 1~cm$^{-1}$, and at 45$^{\circ}$ angle of incidence using a custom-built Fourier transform-based generalized ellipsometer system operating in the polarizer-sample-rotating analyzer arrangement.\cite{Philipp_RSI} The \ac{MIR}-\ac{OHE} measurements were performed at magnetic field strengths of $-$6~T, 0~T and 6~T with the magnetic field vector oriented parallel to the incoming beam. All measurements were performed at room temperature. The optical response of the sample was recorded in terms of $\Psi$ and $\Delta$ for the \ac{MIR}-\ac{SE}, and in terms of Mueller matrix elements for the \ac{MIR}-\ac{OHE} measurements.\cite{Philipp_RSI}

We employed the following optical model for the data analysis: surface-charge-depleted In$_{0.33}$Ga$_{0.67}$N/conductive In$_{0.33}$Ga$_{0.67}$N/AlN/sapphire.
The model dielectric function (MDF) of all constituents contain contribution from IR optical phonon modes, while the surface-charge-depleted and conductive In$_{0.33}$Ga$_{0.67}$N layers contain an additional contribution from free charge carriers. The phonon contribution to the MDF, $\varepsilon_{j}^{L}$, is accounted for by:\cite{Kasic_prb00,schoche_apl13}
\small
\begin{eqnarray}
\label{eqn:lattice}
\varepsilon_{j}^{L} = \varepsilon_{\infty,j}\prod_{l}^{k} \frac{\omega_{\mathrm{LO},l,j}^2 - \omega^2 - i \omega \gamma_{\mathrm{LO},l,j} }{\omega_{\mathrm{TO},l,j}^2 - \omega^2 - i \omega \gamma_{\mathrm{TO},l,j}},
\end{eqnarray}
\normalsize
\noindent
where $\varepsilon_{\infty,j}$ is the high frequency dielectric constant, $k$ denotes the number of active phonon modes for polarization parallel $j = "\parallel"$ (A$_1$ symmetry) and perpendicular $j = "\bot"$ (E$_1$ symmetry) to the $c$-axis; $i$ is the imaginary unit; $\omega$ is the angular frequency; $\omega_{\mathrm{TO},l,j}$, $\omega_{\mathrm{LO},l,j}$ and $\gamma_{\mathrm{TO},l,j}$, $\gamma_{\mathrm{LO},l,j}$ are the \ac{TO} and \ac{LO} modes frequency and broadening parameters, respectively. 

The magnetic field dependent free charge carrier contribution, $\varepsilon^{\mathrm{FCC}}_j$ to the dielectric response of the surface-charge-depleted and conductive In$_{0.33}$Ga$_{0.67}$N layers is accounted for by the classical Drude model augmented by the Lorentz force:\cite{Schubert_OHE,schoche_apl13}

\footnotesize
\begin{eqnarray}
\label{eqn:fcc}
\varepsilon^{\mathrm{FCC}}_j = -\boldsymbol{\omega}_{\mathrm{p}}^2 \begin{pmatrix} \omega^2 \boldsymbol{I} + i\omega\boldsymbol{\gamma} - i\omega \boldsymbol{\omega_{\mathrm{c}}} \begin{pmatrix} 0 & -b_z & b_y\\b_z & 0 & -b_x\\-b_y & b_x & 0 \end{pmatrix} \end{pmatrix}^{-1},
\end{eqnarray}
\normalsize

\noindent
where $\boldsymbol{I}$ is the identity matrix and the magnetic field vector is defined as $\vec{\boldsymbol{B}}=|\vec{\boldsymbol{B}}| (b_{x},b_{y},b_{z})$ in the laboratory coordinate system. At zero magnetic field, the classical Drude contribution is described by the screened plasma frequency tensor, defined as $\boldsymbol{\omega_{\mathrm{p}}}^2 = Nq^2 \boldsymbol{m^{\ast}}^{-1}/(\varepsilon_0 \varepsilon_{\infty})$, and the plasma broadening tensor $\boldsymbol{\gamma} = q \boldsymbol{\mu}^{-1} \boldsymbol{m^{\ast}}^{-1}$, which depend on the free charge carrier concentration parameter $N$, and anisotropic mobility $\boldsymbol{\mu}$ and effective mass $\boldsymbol{m^{\ast}}$ tensors ($\varepsilon_0$ is the vacuum dielectric permittivity, and $q$ the carrier unit charge). At non-zero magnetic field, 
the cyclotron frequency tensor appears, $\boldsymbol{\omega_{\mathrm{c}}} = q |\vec{\boldsymbol{B}}| \boldsymbol{m^{\ast}}^{-1}$, which provides independent access to $\boldsymbol{m^{\ast}}$ and the conductivity type ($q=\pm \mathrm{e}$, where $\mathrm{e}$ is the electron charge and "$+$" corresponds to p-type while "$-$" to n-type conductivity).

The AlN and sapphire MDFs were obtained from previous investigations (Ref.~\onlinecite{schoche_apl13, schubert_prb00}) and not varied during the analysis. The MDFs of the conductive and the surface charge depletion In$_{0.33}$Ga$_{0.67}$N layers shared the same phonon mode contributions with the phonon parameters coupled between the two layers. We found during the \ac{MIR}-\ac{SE} data analysis that a MDF containing one TO-LO phonon mode pair for $j = "\bot"$ (E$_1$ symmetry) and one phonon mode pair $j = "\parallel"$ (A$_1$ symmetry) suffices to excellently match the experimental data. This is in agreement with the one phonon mode behavior previously reported for InGaN A$_1$(LO) phonon mode from Raman scattering experiments.\cite{hernandez_joap05,ager_prb05} Our result on the one phonon mode behavior of the E$_1$(TO) presents the first experimental confirmation of earlier theoretical predictions using the modified random-element isodisplacement and a rigid-ion model.\cite{grille_prb00} The TO and LO mode broadening parameters cannot be differentiated in the case of one phonon mode behavior, and were thus coupled together in our analysis. For  wurtzite III-nitride layers with $c$-plane orientation, \ac{SE} data are not sensitive to the A$_{1}$(TO) phonon parameters.\cite{Kasic_prb00} Thus, in the data analysis the A$_{1}$(TO) frequency was set to 504~cm$^{-1}$ according to a linear interpolation between frequencies for GaN and InN.~\cite{Junqiao} Due to limited sensitivity, the E$_{1}$(LO) phonon frequency $\omega_{\mathrm{LO},\bot}$ was fixed to the linearly interpolated value between the binary parent compounds of 692~cm$^{-1}$.~\cite{Junqiao} We further assumed an isotropic high-frequency dielectric constant ($\varepsilon_{\infty}=\varepsilon_{\infty,\parallel}=\varepsilon_{\infty,\bot}$). The following parameters: $\varepsilon_{\infty}/\varepsilon_{0}$ (5.48), the combined thickness ($d=302$~nm) of the In$_{0.33}$Ga$_{0.67}$N depletion and conductive layers, and the AlN layer thickness ($d_{\mathrm{AlN}}=1024$~nm) were determined from analysis of NIR-UV \ac{SE} measurements and not varied during the \ac{MIR}-\ac{SE} and \ac{OHE} data analysis. The optical model parameters varied during the \ac{MIR}-\ac{SE} and \ac{OHE} data analysis were the remaining phonon and free charge carrier parameters of the conductive and surface depletion In$_{0.33}$Ga$_{0.67}$N layers. The thickness of the surface depletion layer $d_{\mathrm{depl}}$ was varied while the combined In$_{0.33}$Ga$_{0.67}$N layer thickness was held constant. \ac{MIR}-\ac{SE} data ($\Psi$, $\Delta$) and \ac{MIR}-\ac{OHE} difference data ($\Delta$\textbf{M}), obtained by subtracting \ac{OHE} data at 6~T (\textbf{M}(6~T)) and 0~T (\textbf{M}(0~T)), and -6~T (\textbf{M}(-6~T)) magnetic fields, were matched simultaneously.
\begin{figure}[!t]
\centering
\includegraphics[width=.45\textwidth]{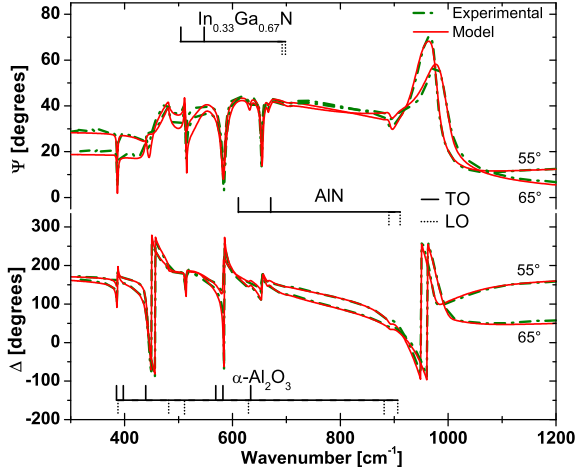}
\caption{
\ac{MIR}-\ac{SE} experimental (solid lines) and the best-match model calculated (dash-dot lines) $\Psi$ and $\Delta$ spectra for the In$_{0.33}$Ga$_{0.67}$N sample. The frequencies of the \ac{IR}-active optical phonons for \ac{TO} (solid lines) and \ac{LO} (dashed lines) in In$_{0.33}$Ga$_{0.67}$N, AlN, and sapphire are indicate by brackets.}
\label{fig:psi_delta}
\end{figure}

Measured and the best-match model calculated \ac{MIR}-\ac{SE} $\Psi$ and $\Delta$ spectra at 55$^{\circ}$ and 65$^{\circ}$ angles of incidence are depicted in Fig.~\ref{fig:psi_delta}. An excellent agreement between measured and calculated spectra is observed. The strong feature at around 950~cm$^{-1}$ is assigned to the A$_{1}$(LO) phonon coupled to free charge carriers in the conductive In$_{0.33}$Ga$_{0.67}$N layer. The subtle spectral feature at around 700~cm$^{-1}$ is associated with the uncoupled A$_{1}$(LO) phonon in the surface depletion layer by virtue of the Berreman effect. A similar behavior was observed and explained for $c$-plane oriented Si-doped GaN, and Mg-doped InN layers.\cite{Kasic_PRB2, schoche_jap13} The broad feature at around 460~cm$^{-1}$ (Fig.~\ref{fig:psi_delta}) can be represented by a weak Lorentz oscillator with polarization perpendicular to the $c$-axis, a small TO-LO splitting and a large damping parameter, assigned as an impurity mode. Such low-polarity modes were also observed previously in doped GaN and InGaN.\cite{Kasic_prb00, Torii_apl03}

\begin{figure}[!t]
\centering
\includegraphics[width=.49\textwidth]{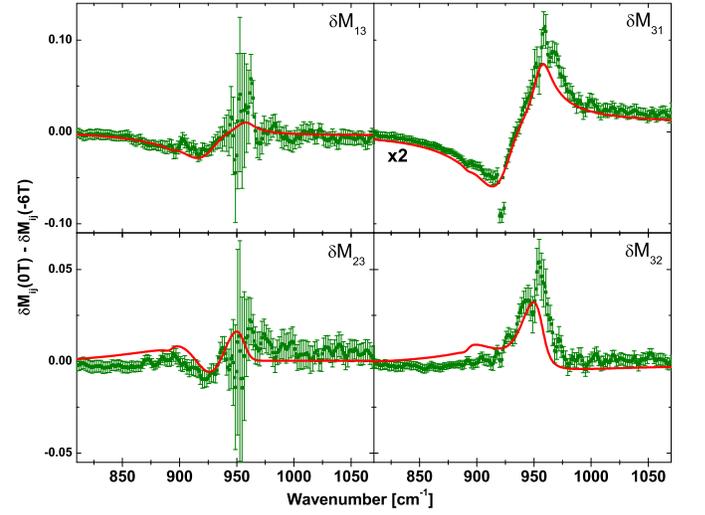}
\caption{
Experimental (dash-dotted lines) and best-match model calculated (solid lines) \ac{MIR}-\ac{OHE} data, shown as differences between data at zero field and at -6T (\textbf{$\Delta$M}($-$6~T)=\textbf{M}(0~T)$-$\textbf{M}($-$6~T)) for elements $\Delta$M$_{ij}$ ($ij=13,31,23,32$) from the In$_{0.33}$Ga$_{0.67}$N sample investigated in this work.
}
\label{fig:diff2}
\end{figure}

Figure~\ref{fig:diff2} depicts experimental and best-match calculated OHE data in terms of Mueller matrix block off-diagonal element ($\Delta$M$_{ij}$, $ij=13,31,23,32$) difference data between \textbf{M}(0~T) and \textbf{M}($-$6~T) data sets. Note that all block off-diagonal Mueller matrix elements are zero without external magnetic field, and all differences therefrom are caused by the magnetic-field induced free charge carrier response in the InGaN layer.

\begin{table}
\caption{Best-match model parameters obtained from \ac{MIR}-\ac{SE} and \ac{MIR}-\ac{OHE} data analysis. Error bars correspond to the 90$\%$ confidence interval.}
\label{tab:table}
\begin{ruledtabular}
\begin{tabular}{c c c c}
Parameter & & Value & \\
\hline
$\omega_{TO,\bot},$						&		& 	$547.3 \pm 0.5$~cm$^{-1}$ 								& \\
$\omega_{LO,\parallel}$				& 	& 	$701 \pm 3$~cm$^{-1}$ 								& \\
$\gamma_{\bot}$ 							& 	& 	$15.9 \pm 0.7$~cm$^{-1}$ 									& \\
$\gamma_{\parallel}$					& 	& 	$18 \pm 8$~cm$^{-1}$ 									& \\
$N$ 													& 	& 	$(1.7 \pm 0.2)\times10^{19}$~cm$^{-3}$ 	& \\
$m^{*}_{\bot}$	 					& 	& 	$(0.205 \pm 0.013)~m_0$										& \\
$m^{*}_{\parallel}$	 			& 	& 	$(0.204 \pm 0.016)~m_0$										& \\
$\mu_{\bot}$ 									& 	& 	$30 \pm 1$~cm$^2/$V${\cdot}$s					& \\
$\mu_{\parallel}$ 						& 	& 	$65 \pm 7$~cm$^2/$V${\cdot}$s					& \\
$N_{\mathrm{depl}}$ 										& 	& 	$<10^{17}$~cm$^{-3}$ 											& \\
$d_{\mathrm{depl}}$ 										& 	& 	$15 \pm 7$~nm													& \\
\end{tabular}
\end{ruledtabular}
\end{table}

The best-match model parameters for the In$_{0.33}$Ga$_{0.67}$N conductive and surface depletion layers are listed in Table~\ref{tab:table}. The electron concentration in the surface depletion layer is found to be below the detection limit of the IR-VASE of 10$^{17}$~cm$^{-3}$. We note that a surface depletion in  In$_{x}$Ga$_{1-x}$N films should occur for In contents below $x \approx 0.4$ (in contrast to alloys with $x \geq 0.4$ where surface accumulation was observed),~\cite{bailey_joap08,king_pssb08} which is consistent with our findings. Our \ac{OHE} results further revealed  n-type conductivity with free electron concentration of $(1.7 \pm 0.2)\times 10^{19}$~cm$^{-3}$ within the conductive In$_{0.33}$Ga$_{0.67}$N layer. In the OHE data analysis the effective mass and mobility parameters were allowed to have different values for polarization perpendicular $m_{\bot}^{\ast}$, $\mu_{\bot}$ and parallel $m_{\parallel}^{\ast}$, $\mu_{\parallel}$ to the $c$-axis. A higher mobility is found along the $c$-axis with $\mu_{\parallel} = 65 \pm 7$~cm$^{2}$/(Vs) and $\mu_{\bot} = 30 \pm 1$~cm$^{2}$/(Vs). A similar anisotropy of the electron mobility parameter was also observed in GaN, InN, and AlGaN alloys and attributed to crystal domains, defect, and impurity distributions in wurtzite-structure, $c$-plane oriented epitaxial layers.\cite{Kasic_prb00, hofmann_jem08, schoche_apl13} The best-match-calculated electron effective mass parameters were determined to be $m^{*}_{\bot}=(0.205 \pm 0.013)~m_0$ and $m^{*}_{\parallel}=(0.204 \pm 0.016)~m_0$. The relative anisotropy of the electron effective mass in InN is $\frac{m^{*}_{\perp}-m^{*}_{||}}{m^{*}_{\perp}+m^{*}_{||}}\approx 15\%$.\cite{hofmann_jem08} Bearing in mind that the electron effective mass in GaN is virtually isotropic,\cite{Kasic_prb00} a slight anisotropy is expected for In$_{0.33}$Ga$_{0.67}$N. However, within  our uncertainty limits, no anisotropy of the electron effective mass is detected. We note that the free electron concentration in the In$_{0.33}$Ga$_{0.67}$N film of 1.7$\times 10^{19}$~cm$^{-3}$ is relatively high, which may play a role for reducing the anisotropy of the effective mass parameter. It was shown that the anisotropy of the electron effective mass parameter decreases with increasing free electron concentration in InN and vanishes for concentrations above 1$\times$10$^{19}$ cm$^{-3}$.\cite{hofmann_jem08} Taking into account the error bars of the determined effective mass parameters, we estimate the highest possible relative anisotropy for In$_{0.33}$Ga$_{0.67}$N to be $\frac{m^{*}_{\perp}-m^{*}_{||}}{m^{*}_{\perp}+m^{*}_{||}}\approx 7\%$. 

\begin{figure}[!t]
\centering
\includegraphics[width=.49\textwidth]{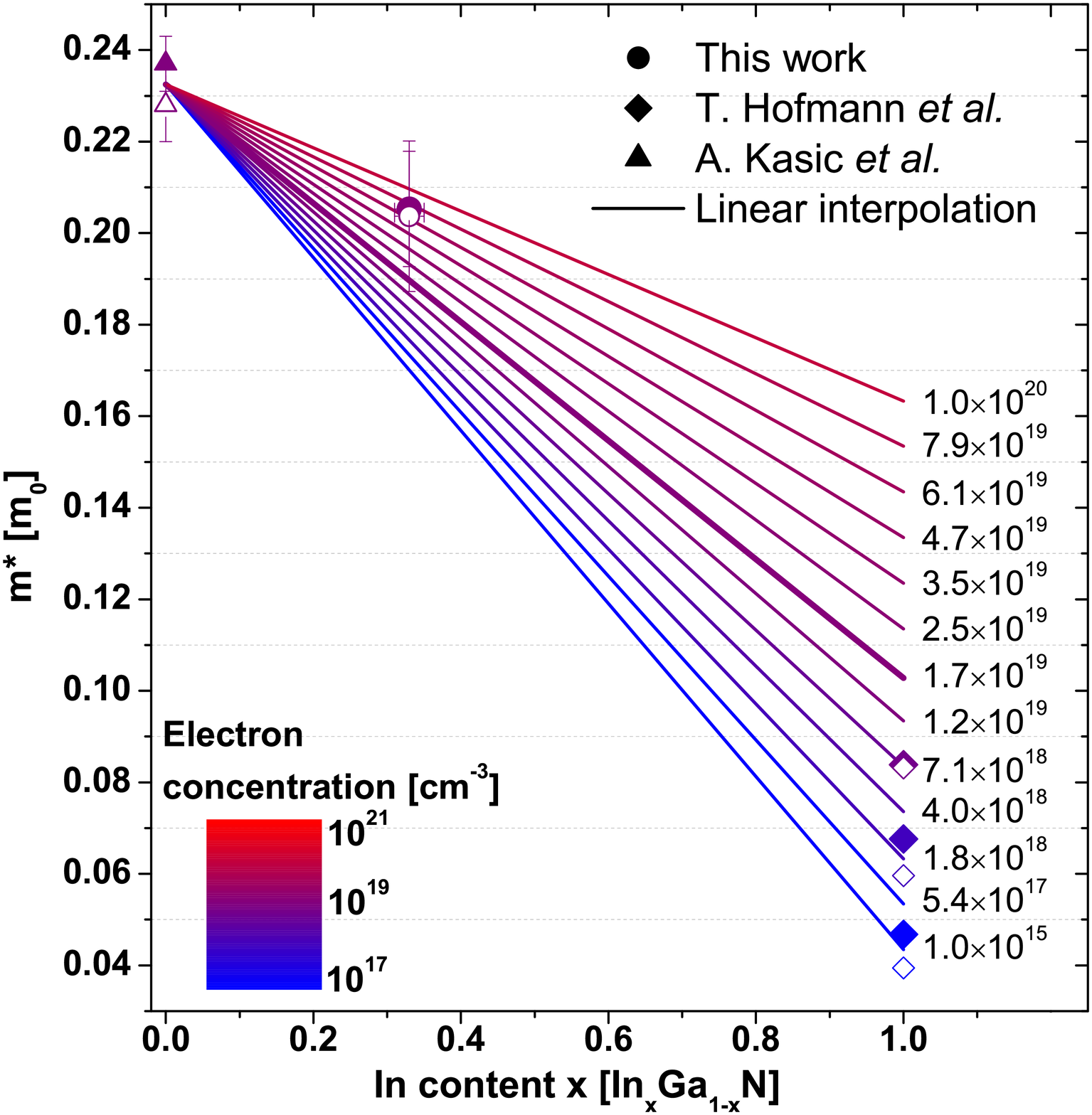}
\caption {Electron effective mass parameter in In$_x$Ga$_{1-x}$N as a function of In content $x$: \ac{MIR}-\ac{OHE} results determined in this work (\ding{108} - $m^{*}_{\bot}$, \ding{109} - $m^{*}_{\parallel}$); InN effective mass from Ref.~\onlinecite{hofmann_jem08} ($\Diamondblack$ - $m^{*}_{\bot}$, $\Diamond$ - $m^{*}_{\parallel}$) and GaN effective mass from Ref.~\onlinecite{Kasic_prb00} (\ding{115} - $m^{*}_{\bot}$, $\bigtriangleup$ - $m^{*}_{\parallel}$). The solid lines indicate the linear dependences of the isotropically averaged In$_x$Ga$_{1-x}$N electron effective mass parameters on In content for different electron concentrations obtained from a linear interpolation of the results in Refs.~\onlinecite{hofmann_jem08} (InN) and Refs.~\onlinecite{Kasic_prb00} (GaN), considering the nonparaobolicity in InN, and assuming full parabolic behavior in GaN. The color coding represents the free electron concentrations in In$_x$Ga$_{1-x}$N.}
\label{fig:effmass}
\end{figure}

Figure~\ref{fig:effmass} summarizes our results together with previously reported effective mass parameters for InN~\cite{hofmann_jem08} and GaN.~\cite{Kasic_prb00} The solid lines indicate the linear interpolation between the isotropically averaged electron effective mass parameters of GaN and InN for different electron concentrations, assuming parabolic bands in GaN and considering the nonparabolicity in InN.~\cite{Fu_1,hofmann_jem08} The GaN value reported by Kasic~\textit{et al.}~\cite{Kasic_prb00} is considered to be independent on the electron concentration.~\cite{Feneberg_1} The linear interpolation for the lowest carrier concentration of $10^{15}$~cm$^{-3}$ may be seen as the curvature of the conduction band bottom in In$_x$Ga$_{1-x}$N. The electron effective mass parameter determined in this work is larger than the value at the bottom of the conduction band  predicted by the linear interpolation. In principle, this may be due to two effects: i) a nonparabolic  conduction band in In$_{0.33}$Ga$_{0.67}$N leading to dependence of the electron effective mass on carrier concentration, and/or ii) a nonlinear dependence  $m^*(x)$ with significant bowing parameter. A very small effective mass bowing parameter was theoretically predicted for In$_{x}$Ga$_{1-x}$N.\cite{Elfitur} We therefore compare our effective mass parameters with the linear interpolation for the electron concentration in our sample of $1.7\times 10^{19}$~cm$^{-3}$. It can be seen from Fig.~\ref{fig:effmass} that within the error bar our result is consistent with this linear interpolation, and thus indicative for nonparabolicity in In$_{0.33}$Ga$_{0.67}$N. A larger bowing parameter, however, cannot be ruled out. Further accurate effective mass parameters in In$_x$Ga$_{1-x}$N  with different electron concentrations and different In contents are needed.

In conclusion, the room temperature electron effective mass parameters perpendicular and parallel to the the $c$-axis of In$_{0.33}$Ga$_{0.67}$N epitaxial layer are determined to be $m^{*}_{\bot}=(0.205 \pm 0.013)~m_0$ and $m^{*}_{\parallel}=(0.204 \pm 0.016)~m_0$ for a free electron concentration of $(1.7 \pm 0.2)\times 10^{19}$~cm$^{-3}$. Within our uncertainty
limits, the electron effective mass parameter is found to be isotropic and the upper limit of the possible relative anisotropy is estimated to be 7\%. Our results indicate that nonparabolicity of the conduction band in In$_{0.33}$Ga$_{0.67}$N is likely present. The free electron mobility is found to be anisotropic with $\mu_{\bot} = 30 \pm 1$~cm$^{2}$/(Vs) and $\mu_{\parallel} = 65 \pm 7$~cm$^{2}$/(Vs). Our results confirm previous experimental and theoretical findings for the one mode behavior of the A$_1$(LO) phonon and provide experimental support for the theoretical predictions of a one mode behavior of the E$_1$(TO) phonon in InGaN.

This work was supported by the Swedish Research Council (VR) under Grant No. 2013-5580 and 2016-00889, the Swedish Governmental Agency for Innovation Systems (VINNOVA) under the VINNMER international qualification program, Grant No. 2011-03486, the Swedish Government Strategic Research Area in Materials Science on Functional Materials at Link\"oping University, Faculty Grant SFO Mat LiU No. 2009 00971, and the Swedish Foundation for Strategic Research (SSF), under Grant No. FL12-0181 and RIF14-055. The authors further acknowledge financial support by the University of Nebraska-Lincoln, the J.~A.~Woollam Co., Inc., the J.~A.~Woollam Foundation and the National Science Foundation (awards MRSEC DMR 1420645, CMMI 1337856 and EAR 1521428). The authors thank Dr.~Mengyao~Xie and Prof.~Dr.~Enrique~Calleja (Universidad Polit\'{e}cnica de Madrid) for providing the In$_{0.33}$Ga$_{0.67}$N sample and Laurent Souqui (Link\"{o}ping University) for assistance with XRD measurements.
%\end{acknowledgments}

\bibliography{bib}

\begin{thebibliography}{24}%
\makeatletter
\providecommand \@ifxundefined [1]{%
 \@ifx{#1\undefined}
}%
\providecommand \@ifnum [1]{%
 \ifnum #1\expandafter \@firstoftwo
 \else \expandafter \@secondoftwo
 \fi
}%
\providecommand \@ifx [1]{%
 \ifx #1\expandafter \@firstoftwo
 \else \expandafter \@secondoftwo
 \fi
}%
\providecommand \natexlab [1]{#1}%
\providecommand \enquote  [1]{``#1''}%
\providecommand \bibnamefont  [1]{#1}%
\providecommand \bibfnamefont [1]{#1}%
\providecommand \citenamefont [1]{#1}%
\providecommand \href@noop [0]{\@secondoftwo}%
\providecommand \href [0]{\begingroup \@sanitize@url \@href}%
\providecommand \@href[1]{\@@startlink{#1}\@@href}%
\providecommand \@@href[1]{\endgroup#1\@@endlink}%
\providecommand \@sanitize@url [0]{\catcode `\\12\catcode `\$12\catcode
  `\&12\catcode `\#12\catcode `\^12\catcode `\_12\catcode `\%12\relax}%
\providecommand \@@startlink[1]{}%
\providecommand \@@endlink[0]{}%
\providecommand \url  [0]{\begingroup\@sanitize@url \@url }%
\providecommand \@url [1]{\endgroup\@href {#1}{\urlprefix }}%
\providecommand \urlprefix  [0]{URL }%
\providecommand \Eprint [0]{\href }%
\providecommand \doibase [0]{http://dx.doi.org/}%
\providecommand \selectlanguage [0]{\@gobble}%
\providecommand \bibinfo  [0]{\@secondoftwo}%
\providecommand \bibfield  [0]{\@secondoftwo}%
\providecommand \translation [1]{[#1]}%
\providecommand \BibitemOpen [0]{}%
\providecommand \bibitemStop [0]{}%
\providecommand \bibitemNoStop [0]{.\EOS\space}%
\providecommand \EOS [0]{\spacefactor3000\relax}%
\providecommand \BibitemShut  [1]{\csname bibitem#1\endcsname}%
\let\auto@bib@innerbib\@empty
%</preamble>
\bibitem [{\citenamefont {Wu}(2009)}]{Junqiao}%
  \BibitemOpen
  \bibfield  {author} {\bibinfo {author} {\bibfnamefont {J.}~\bibnamefont
  {Wu}},\ }\href {\doibase 10.1063/1.3155798} {\bibfield  {journal} {\bibinfo
  {journal} {J. Appl. Phys.}\ }\textbf {\bibinfo {volume} {106}},\ \bibinfo
  {pages} {011101} (\bibinfo {year} {2009})}\BibitemShut {NoStop}%
\bibitem [{\citenamefont {Ardaravi\v{c}ius}\ \emph {et~al.}(2015)\citenamefont
  {Ardaravi\v{c}ius}, \citenamefont {Kiprijanovi\v{c}}, \citenamefont
  {Liberis}, \citenamefont {\v{S}ermuk\v{s}nis}, \citenamefont {Matulionis},
  \citenamefont {Ferreyra}, \citenamefont {Avrutin}, \citenamefont
  {\"{O}zg\"{u}r},\ and\ \citenamefont {Morko\c{c}}}]{ardaravicius_ssat15}%
  \BibitemOpen
  \bibfield  {author} {\bibinfo {author} {\bibfnamefont {L.}~\bibnamefont
  {Ardaravi\v{c}ius}}, \bibinfo {author} {\bibfnamefont {O.}~\bibnamefont
  {Kiprijanovi\v{c}}}, \bibinfo {author} {\bibfnamefont {J.}~\bibnamefont
  {Liberis}}, \bibinfo {author} {\bibfnamefont {E.}~\bibnamefont
  {\v{S}ermuk\v{s}nis}}, \bibinfo {author} {\bibfnamefont {A.}~\bibnamefont
  {Matulionis}}, \bibinfo {author} {\bibfnamefont {R.~A.}\ \bibnamefont
  {Ferreyra}}, \bibinfo {author} {\bibfnamefont {V.}~\bibnamefont {Avrutin}},
  \bibinfo {author} {\bibfnamefont {U.}~\bibnamefont {\"{O}zg\"{u}r}}, \ and\
  \bibinfo {author} {\bibfnamefont {H.}~\bibnamefont {Morko\c{c}}},\ }\href
  {http://stacks.iop.org/0268-1242/30/i=10/a=105016} {\bibfield  {journal}
  {\bibinfo  {journal} {Semicond. Sci. Technol.}\ }\textbf {\bibinfo {volume}
  {30}},\ \bibinfo {pages} {105016} (\bibinfo {year} {2015})}\BibitemShut
  {NoStop}%
\bibitem [{\citenamefont {Elfituri}\ and\ \citenamefont
  {Hourahine}(2012)}]{Elfitur}%
  \BibitemOpen
  \bibfield  {author} {\bibinfo {author} {\bibfnamefont {F.}~\bibnamefont
  {Elfituri}}\ and\ \bibinfo {author} {\bibfnamefont {B.}~\bibnamefont
  {Hourahine}},\ }\href {\doibase 10.1002/pssa.201100147} {\bibfield  {journal}
  {\bibinfo  {journal} {Phys. Status Solidi A}\ }\textbf {\bibinfo {volume}
  {209}},\ \bibinfo {pages} {79} (\bibinfo {year} {2012})}\BibitemShut
  {NoStop}%
\bibitem [{\citenamefont {Millot}\ \emph {et~al.}(2011)\citenamefont {Millot},
  \citenamefont {Ubrig}, \citenamefont {Poumirol}, \citenamefont {Gherasoiu},
  \citenamefont {Walukiewicz}, \citenamefont {George}, \citenamefont
  {Portugall}, \citenamefont {L\'eotin}, \citenamefont {Goiran},\ and\
  \citenamefont {Broto}}]{millot_prb11}%
  \BibitemOpen
  \bibfield  {author} {\bibinfo {author} {\bibfnamefont {M.}~\bibnamefont
  {Millot}}, \bibinfo {author} {\bibfnamefont {N.}~\bibnamefont {Ubrig}},
  \bibinfo {author} {\bibfnamefont {J.~M.}\ \bibnamefont {Poumirol}}, \bibinfo
  {author} {\bibfnamefont {I.}~\bibnamefont {Gherasoiu}}, \bibinfo {author}
  {\bibfnamefont {W.}~\bibnamefont {Walukiewicz}}, \bibinfo {author}
  {\bibfnamefont {S.}~\bibnamefont {George}}, \bibinfo {author} {\bibfnamefont
  {O.}~\bibnamefont {Portugall}}, \bibinfo {author} {\bibfnamefont
  {J.}~\bibnamefont {L\'eotin}}, \bibinfo {author} {\bibfnamefont
  {M.}~\bibnamefont {Goiran}}, \ and\ \bibinfo {author} {\bibfnamefont {J.~M.}\
  \bibnamefont {Broto}},\ }\href {\doibase 10.1103/PhysRevB.83.125204}
  {\bibfield  {journal} {\bibinfo  {journal} {Phys. Rev. B}\ }\textbf {\bibinfo
  {volume} {83}},\ \bibinfo {pages} {125204} (\bibinfo {year}
  {2011})}\BibitemShut {NoStop}%
\bibitem [{\citenamefont {Yadav}\ \emph {et~al.}(2014)\citenamefont {Yadav},
  \citenamefont {Mohanta}, \citenamefont {Srinivasa},\ and\ \citenamefont
  {Major}}]{Yadav}%
  \BibitemOpen
  \bibfield  {author} {\bibinfo {author} {\bibfnamefont {B.~S.}\ \bibnamefont
  {Yadav}}, \bibinfo {author} {\bibfnamefont {P.}~\bibnamefont {Mohanta}},
  \bibinfo {author} {\bibfnamefont {R.}~\bibnamefont {Srinivasa}}, \ and\
  \bibinfo {author} {\bibfnamefont {S.}~\bibnamefont {Major}},\ }\href
  {\doibase http://dx.doi.org/10.1016/j.tsf.2013.11.117} {\bibfield  {journal}
  {\bibinfo  {journal} {Thin Solid Films}\ }\textbf {\bibinfo {volume} {555}},\
  \bibinfo {pages} {179 } (\bibinfo {year} {2014})}\BibitemShut {NoStop}%
\bibitem [{\citenamefont {Eljarrat}\ \emph {et~al.}(2016)\citenamefont
  {Eljarrat}, \citenamefont {Lopez-Conesa}, \citenamefont {Magen},
  \citenamefont {Garcia-Lepetit}, \citenamefont {Gacevic}, \citenamefont
  {Calleja}, \citenamefont {Peiro},\ and\ \citenamefont {Estrade}}]{Eljarrat}%
  \BibitemOpen
  \bibfield  {author} {\bibinfo {author} {\bibfnamefont {A.}~\bibnamefont
  {Eljarrat}}, \bibinfo {author} {\bibfnamefont {L.}~\bibnamefont
  {Lopez-Conesa}}, \bibinfo {author} {\bibfnamefont {C.}~\bibnamefont {Magen}},
  \bibinfo {author} {\bibfnamefont {N.}~\bibnamefont {Garcia-Lepetit}},
  \bibinfo {author} {\bibfnamefont {Z.}~\bibnamefont {Gacevic}}, \bibinfo
  {author} {\bibfnamefont {E.}~\bibnamefont {Calleja}}, \bibinfo {author}
  {\bibfnamefont {F.}~\bibnamefont {Peiro}}, \ and\ \bibinfo {author}
  {\bibfnamefont {S.}~\bibnamefont {Estrade}},\ }\href {\doibase
  10.1039/C6CP04493J} {\bibfield  {journal} {\bibinfo  {journal} {Phys. Chem.
  Chem. Phys.}\ }\textbf {\bibinfo {volume} {18}},\ \bibinfo {pages} {23264}
  (\bibinfo {year} {2016})}\BibitemShut {NoStop}%
\bibitem [{\citenamefont {Walukiewicz}\ \emph {et~al.}(2006)\citenamefont
  {Walukiewicz}, \citenamefont {III}, \citenamefont {Yu}, \citenamefont
  {Liliental-Weber}, \citenamefont {Wu}, \citenamefont {Li}, \citenamefont
  {Jones},\ and\ \citenamefont {Denlinger}}]{Walukiewicz1}%
  \BibitemOpen
  \bibfield  {author} {\bibinfo {author} {\bibfnamefont {W.}~\bibnamefont
  {Walukiewicz}}, \bibinfo {author} {\bibfnamefont {J.~W.~A.}\ \bibnamefont
  {III}}, \bibinfo {author} {\bibfnamefont {K.~M.}\ \bibnamefont {Yu}},
  \bibinfo {author} {\bibfnamefont {Z.}~\bibnamefont {Liliental-Weber}},
  \bibinfo {author} {\bibfnamefont {J.}~\bibnamefont {Wu}}, \bibinfo {author}
  {\bibfnamefont {S.~X.}\ \bibnamefont {Li}}, \bibinfo {author} {\bibfnamefont
  {R.~E.}\ \bibnamefont {Jones}}, \ and\ \bibinfo {author} {\bibfnamefont
  {J.~D.}\ \bibnamefont {Denlinger}},\ }\href
  {http://stacks.iop.org/0022-3727/39/i=5/a=R01} {\bibfield  {journal}
  {\bibinfo  {journal} {J. Phys. D: Appl. Phys.}\ }\textbf {\bibinfo {volume}
  {39}},\ \bibinfo {pages} {R83} (\bibinfo {year} {2006})}\BibitemShut
  {NoStop}%
\bibitem [{\citenamefont {Hofmann}\ \emph {et~al.}(2008)\citenamefont
  {Hofmann}, \citenamefont {Darakchieva}, \citenamefont {Monemar},
  \citenamefont {Lu}, \citenamefont {Schaff},\ and\ \citenamefont
  {Schubert}}]{hofmann_jem08}%
  \BibitemOpen
  \bibfield  {author} {\bibinfo {author} {\bibfnamefont {T.}~\bibnamefont
  {Hofmann}}, \bibinfo {author} {\bibfnamefont {V.}~\bibnamefont
  {Darakchieva}}, \bibinfo {author} {\bibfnamefont {B.}~\bibnamefont
  {Monemar}}, \bibinfo {author} {\bibfnamefont {H.}~\bibnamefont {Lu}},
  \bibinfo {author} {\bibfnamefont {W.}~\bibnamefont {Schaff}}, \ and\ \bibinfo
  {author} {\bibfnamefont {M.}~\bibnamefont {Schubert}},\ }\href {\doibase
  10.1007/s11664-008-0385-8} {\bibfield  {journal} {\bibinfo  {journal} {J.
  Electron. Mater.}\ }\textbf {\bibinfo {volume} {37}},\ \bibinfo {pages} {611}
  (\bibinfo {year} {2008})}\BibitemShut {NoStop}%
\bibitem [{\citenamefont {Kasic}\ \emph {et~al.}(2000)\citenamefont {Kasic},
  \citenamefont {Schubert}, \citenamefont {Einfeldt}, \citenamefont {Hommel},\
  and\ \citenamefont {Tiwald}}]{Kasic_prb00}%
  \BibitemOpen
  \bibfield  {author} {\bibinfo {author} {\bibfnamefont {A.}~\bibnamefont
  {Kasic}}, \bibinfo {author} {\bibfnamefont {M.}~\bibnamefont {Schubert}},
  \bibinfo {author} {\bibfnamefont {S.}~\bibnamefont {Einfeldt}}, \bibinfo
  {author} {\bibfnamefont {D.}~\bibnamefont {Hommel}}, \ and\ \bibinfo {author}
  {\bibfnamefont {T.~E.}\ \bibnamefont {Tiwald}},\ }\href {\doibase
  10.1103/PhysRevB.62.7365} {\bibfield  {journal} {\bibinfo  {journal} {Phys.
  Rev. B}\ }\textbf {\bibinfo {volume} {62}},\ \bibinfo {pages} {7365}
  (\bibinfo {year} {2000})}\BibitemShut {NoStop}%
\bibitem [{\citenamefont {Feneberg}\ \emph {et~al.}(2013)\citenamefont
  {Feneberg}, \citenamefont {Lange}, \citenamefont {Lidig}, \citenamefont
  {Wieneke}, \citenamefont {Witte}, \citenamefont {Bl\"{a}sing}, \citenamefont
  {Dadgar}, \citenamefont {Krost},\ and\ \citenamefont
  {Goldhahn}}]{Feneberg_1}%
  \BibitemOpen
  \bibfield  {author} {\bibinfo {author} {\bibfnamefont {M.}~\bibnamefont
  {Feneberg}}, \bibinfo {author} {\bibfnamefont {K.}~\bibnamefont {Lange}},
  \bibinfo {author} {\bibfnamefont {C.}~\bibnamefont {Lidig}}, \bibinfo
  {author} {\bibfnamefont {M.}~\bibnamefont {Wieneke}}, \bibinfo {author}
  {\bibfnamefont {H.}~\bibnamefont {Witte}}, \bibinfo {author} {\bibfnamefont
  {J.}~\bibnamefont {Bl\"{a}sing}}, \bibinfo {author} {\bibfnamefont
  {A.}~\bibnamefont {Dadgar}}, \bibinfo {author} {\bibfnamefont
  {A.}~\bibnamefont {Krost}}, \ and\ \bibinfo {author} {\bibfnamefont
  {R.}~\bibnamefont {Goldhahn}},\ }\href {\doibase
  http://dx.doi.org/10.1063/1.4840055} {\bibfield  {journal} {\bibinfo
  {journal} {Appl. Phys. Lett.}\ }\textbf {\bibinfo {volume} {103}},\ \bibinfo
  {eid} {232104} (\bibinfo {year} {2013}),\
  http://dx.doi.org/10.1063/1.4840055}\BibitemShut {NoStop}%
\bibitem [{\citenamefont {Schubert}\ \emph {et~al.}(2003)\citenamefont
  {Schubert}, \citenamefont {Hofmann},\ and\ \citenamefont
  {Herzinger}}]{Schubert:03}%
  \BibitemOpen
  \bibfield  {author} {\bibinfo {author} {\bibfnamefont {M.}~\bibnamefont
  {Schubert}}, \bibinfo {author} {\bibfnamefont {T.}~\bibnamefont {Hofmann}}, \
  and\ \bibinfo {author} {\bibfnamefont {C.~M.}\ \bibnamefont {Herzinger}},\
  }\href {\doibase 10.1364/JOSAA.20.000347} {\bibfield  {journal} {\bibinfo
  {journal} {J. Opt. Soc. Am. A}\ }\textbf {\bibinfo {volume} {20}},\ \bibinfo
  {pages} {347} (\bibinfo {year} {2003})}\BibitemShut {NoStop}%
\bibitem [{\citenamefont {Schubert}\ \emph {et~al.}(2016)\citenamefont
  {Schubert}, \citenamefont {K\"{u}hne}, \citenamefont {Darakchieva},\ and\
  \citenamefont {Hofmann}}]{Schubert_OHE}%
  \BibitemOpen
  \bibfield  {author} {\bibinfo {author} {\bibfnamefont {M.}~\bibnamefont
  {Schubert}}, \bibinfo {author} {\bibfnamefont {P.}~\bibnamefont {K\"{u}hne}},
  \bibinfo {author} {\bibfnamefont {V.}~\bibnamefont {Darakchieva}}, \ and\
  \bibinfo {author} {\bibfnamefont {T.}~\bibnamefont {Hofmann}},\ }\href
  {\doibase 10.1364/JOSAA.33.001553} {\bibfield  {journal} {\bibinfo  {journal}
  {J. Opt. Soc. Am. A}\ }\textbf {\bibinfo {volume} {33}},\ \bibinfo {pages}
  {1553} (\bibinfo {year} {2016})}\BibitemShut {NoStop}%
\bibitem [{\citenamefont {K\"{u}hne}\ \emph {et~al.}(2014)\citenamefont
  {K\"{u}hne}, \citenamefont {Herzinger}, \citenamefont {Schubert},
  \citenamefont {Woollam},\ and\ \citenamefont {Hofmann}}]{Philipp_RSI}%
  \BibitemOpen
  \bibfield  {author} {\bibinfo {author} {\bibfnamefont {P.}~\bibnamefont
  {K\"{u}hne}}, \bibinfo {author} {\bibfnamefont {C.~M.}\ \bibnamefont
  {Herzinger}}, \bibinfo {author} {\bibfnamefont {M.}~\bibnamefont {Schubert}},
  \bibinfo {author} {\bibfnamefont {J.~A.}\ \bibnamefont {Woollam}}, \ and\
  \bibinfo {author} {\bibfnamefont {T.}~\bibnamefont {Hofmann}},\ }\href
  {\doibase http://dx.doi.org/10.1063/1.4889920} {\bibfield  {journal}
  {\bibinfo  {journal} {Rev. Sci. Instrum.}\ }\textbf {\bibinfo {volume}
  {85}},\ \bibinfo {eid} {071301} (\bibinfo {year} {2014})}\BibitemShut
  {NoStop}%
\bibitem [{\citenamefont {Sch\"{o}che}\ \emph
  {et~al.}(2013{\natexlab{a}})\citenamefont {Sch\"{o}che}, \citenamefont
  {K\"{u}hne}, \citenamefont {Hofmann}, \citenamefont {Schubert}, \citenamefont
  {Nilsson}, \citenamefont {Kakanakova-Georgieva}, \citenamefont {Janz{\'e}n},\
  and\ \citenamefont {Darakchieva}}]{schoche_apl13}%
  \BibitemOpen
  \bibfield  {author} {\bibinfo {author} {\bibfnamefont {S.}~\bibnamefont
  {Sch\"{o}che}}, \bibinfo {author} {\bibfnamefont {P.}~\bibnamefont
  {K\"{u}hne}}, \bibinfo {author} {\bibfnamefont {T.}~\bibnamefont {Hofmann}},
  \bibinfo {author} {\bibfnamefont {M.}~\bibnamefont {Schubert}}, \bibinfo
  {author} {\bibfnamefont {D.}~\bibnamefont {Nilsson}}, \bibinfo {author}
  {\bibfnamefont {A.}~\bibnamefont {Kakanakova-Georgieva}}, \bibinfo {author}
  {\bibfnamefont {E.}~\bibnamefont {Janz{\'e}n}}, \ and\ \bibinfo {author}
  {\bibfnamefont {V.}~\bibnamefont {Darakchieva}},\ }\href {\doibase
  10.1063/1.4833195} {\bibfield  {journal} {\bibinfo  {journal} {Appl. Phys.
  Lett.}\ }\textbf {\bibinfo {volume} {103}},\ \bibinfo {pages} {212107}
  (\bibinfo {year} {2013}{\natexlab{a}})}\BibitemShut {NoStop}%
\bibitem [{\citenamefont {Schubert}\ \emph {et~al.}(2000)\citenamefont
  {Schubert}, \citenamefont {Tiwald},\ and\ \citenamefont
  {Herzinger}}]{schubert_prb00}%
  \BibitemOpen
  \bibfield  {author} {\bibinfo {author} {\bibfnamefont {M.}~\bibnamefont
  {Schubert}}, \bibinfo {author} {\bibfnamefont {T.~E.}\ \bibnamefont
  {Tiwald}}, \ and\ \bibinfo {author} {\bibfnamefont {C.~M.}\ \bibnamefont
  {Herzinger}},\ }\href {\doibase 10.1103/PhysRevB.61.8187} {\bibfield
  {journal} {\bibinfo  {journal} {Phys. Rev. B}\ }\textbf {\bibinfo {volume}
  {61}},\ \bibinfo {pages} {8187} (\bibinfo {year} {2000})}\BibitemShut
  {NoStop}%
\bibitem [{\citenamefont {Hern{\'a}ndez}\ \emph {et~al.}(2005)\citenamefont
  {Hern{\'a}ndez}, \citenamefont {Cusc{\'o}}, \citenamefont {Pastor},
  \citenamefont {Art{\'u}s}, \citenamefont {O’Donnell}, \citenamefont
  {Martin}, \citenamefont {Watson}, \citenamefont {Nanishi},\ and\
  \citenamefont {Calleja}}]{hernandez_joap05}%
  \BibitemOpen
  \bibfield  {author} {\bibinfo {author} {\bibfnamefont {S.}~\bibnamefont
  {Hern{\'a}ndez}}, \bibinfo {author} {\bibfnamefont {R.}~\bibnamefont
  {Cusc{\'o}}}, \bibinfo {author} {\bibfnamefont {D.}~\bibnamefont {Pastor}},
  \bibinfo {author} {\bibfnamefont {L.}~\bibnamefont {Art{\'u}s}}, \bibinfo
  {author} {\bibfnamefont {K.}~\bibnamefont {O’Donnell}}, \bibinfo {author}
  {\bibfnamefont {R.}~\bibnamefont {Martin}}, \bibinfo {author} {\bibfnamefont
  {I.}~\bibnamefont {Watson}}, \bibinfo {author} {\bibfnamefont
  {Y.}~\bibnamefont {Nanishi}}, \ and\ \bibinfo {author} {\bibfnamefont
  {E.}~\bibnamefont {Calleja}},\ }\href@noop {} {\bibfield  {journal} {\bibinfo
   {journal} {J. Appl. Phys.}\ }\textbf {\bibinfo {volume} {98}},\ \bibinfo
  {pages} {013511} (\bibinfo {year} {2005})}\BibitemShut {NoStop}%
\bibitem [{\citenamefont {Ager~III}\ \emph {et~al.}(2005)\citenamefont
  {Ager~III}, \citenamefont {Walukiewicz}, \citenamefont {Shan}, \citenamefont
  {Yu}, \citenamefont {Li}, \citenamefont {Haller}, \citenamefont {Lu},\ and\
  \citenamefont {Schaff}}]{ager_prb05}%
  \BibitemOpen
  \bibfield  {author} {\bibinfo {author} {\bibfnamefont {J.}~\bibnamefont
  {Ager~III}}, \bibinfo {author} {\bibfnamefont {W.}~\bibnamefont
  {Walukiewicz}}, \bibinfo {author} {\bibfnamefont {W.}~\bibnamefont {Shan}},
  \bibinfo {author} {\bibfnamefont {K.}~\bibnamefont {Yu}}, \bibinfo {author}
  {\bibfnamefont {S.}~\bibnamefont {Li}}, \bibinfo {author} {\bibfnamefont
  {E.}~\bibnamefont {Haller}}, \bibinfo {author} {\bibfnamefont
  {H.}~\bibnamefont {Lu}}, \ and\ \bibinfo {author} {\bibfnamefont
  {W.}~\bibnamefont {Schaff}},\ }\href@noop {} {\bibfield  {journal} {\bibinfo
  {journal} {Phys. Rev. B}\ }\textbf {\bibinfo {volume} {72}},\ \bibinfo
  {pages} {155204} (\bibinfo {year} {2005})}\BibitemShut {NoStop}%
\bibitem [{\citenamefont {Grille}\ \emph {et~al.}(2000)\citenamefont {Grille},
  \citenamefont {Schnittler},\ and\ \citenamefont {Bechstedt}}]{grille_prb00}%
  \BibitemOpen
  \bibfield  {author} {\bibinfo {author} {\bibfnamefont {H.}~\bibnamefont
  {Grille}}, \bibinfo {author} {\bibfnamefont {C.}~\bibnamefont {Schnittler}},
  \ and\ \bibinfo {author} {\bibfnamefont {F.}~\bibnamefont {Bechstedt}},\
  }\href@noop {} {\bibfield  {journal} {\bibinfo  {journal} {Phys. Rev. B}\
  }\textbf {\bibinfo {volume} {61}},\ \bibinfo {pages} {6091} (\bibinfo {year}
  {2000})}\BibitemShut {NoStop}%
\bibitem [{\citenamefont {Kasic}\ \emph {et~al.}(2002)\citenamefont {Kasic},
  \citenamefont {Schubert}, \citenamefont {Saito}, \citenamefont {Nanishi},\
  and\ \citenamefont {Wagner}}]{Kasic_PRB2}%
  \BibitemOpen
  \bibfield  {author} {\bibinfo {author} {\bibfnamefont {A.}~\bibnamefont
  {Kasic}}, \bibinfo {author} {\bibfnamefont {M.}~\bibnamefont {Schubert}},
  \bibinfo {author} {\bibfnamefont {Y.}~\bibnamefont {Saito}}, \bibinfo
  {author} {\bibfnamefont {Y.}~\bibnamefont {Nanishi}}, \ and\ \bibinfo
  {author} {\bibfnamefont {G.}~\bibnamefont {Wagner}},\ }\href {\doibase
  10.1103/PhysRevB.65.115206} {\bibfield  {journal} {\bibinfo  {journal} {Phys.
  Rev. B}\ }\textbf {\bibinfo {volume} {65}},\ \bibinfo {pages} {115206}
  (\bibinfo {year} {2002})}\BibitemShut {NoStop}%
\bibitem [{\citenamefont {Sch\"{o}che}\ \emph
  {et~al.}(2013{\natexlab{b}})\citenamefont {Sch\"{o}che}, \citenamefont
  {Hofmann}, \citenamefont {Darakchieva}, \citenamefont {Sedrine},
  \citenamefont {Wang}, \citenamefont {Yoshikawa},\ and\ \citenamefont
  {Schubert}}]{schoche_jap13}%
  \BibitemOpen
  \bibfield  {author} {\bibinfo {author} {\bibfnamefont {S.}~\bibnamefont
  {Sch\"{o}che}}, \bibinfo {author} {\bibfnamefont {T.}~\bibnamefont
  {Hofmann}}, \bibinfo {author} {\bibfnamefont {V.}~\bibnamefont
  {Darakchieva}}, \bibinfo {author} {\bibfnamefont {N.~B.}\ \bibnamefont
  {Sedrine}}, \bibinfo {author} {\bibfnamefont {X.}~\bibnamefont {Wang}},
  \bibinfo {author} {\bibfnamefont {A.}~\bibnamefont {Yoshikawa}}, \ and\
  \bibinfo {author} {\bibfnamefont {M.}~\bibnamefont {Schubert}},\ }\href
  {\doibase 10.1063/1.4772625} {\bibfield  {journal} {\bibinfo  {journal} {J.
  Appl. Phys.}\ }\textbf {\bibinfo {volume} {113}},\ \bibinfo {pages} {013502}
  (\bibinfo {year} {2013}{\natexlab{b}})}\BibitemShut {NoStop}%
\bibitem [{\citenamefont {Torii}\ \emph {et~al.}(2003)\citenamefont {Torii},
  \citenamefont {Usukura}, \citenamefont {Nakamura}, \citenamefont {Sota},
  \citenamefont {Chichibu}, \citenamefont {Kitamura},\ and\ \citenamefont
  {Okumura}}]{Torii_apl03}%
  \BibitemOpen
  \bibfield  {author} {\bibinfo {author} {\bibfnamefont {K.}~\bibnamefont
  {Torii}}, \bibinfo {author} {\bibfnamefont {N.}~\bibnamefont {Usukura}},
  \bibinfo {author} {\bibfnamefont {A.}~\bibnamefont {Nakamura}}, \bibinfo
  {author} {\bibfnamefont {T.}~\bibnamefont {Sota}}, \bibinfo {author}
  {\bibfnamefont {S.~F.}\ \bibnamefont {Chichibu}}, \bibinfo {author}
  {\bibfnamefont {T.}~\bibnamefont {Kitamura}}, \ and\ \bibinfo {author}
  {\bibfnamefont {H.}~\bibnamefont {Okumura}},\ }\href {\doibase
  10.1063/1.1535273} {\bibfield  {journal} {\bibinfo  {journal} {Appl. Phys.
  Lett.}\ }\textbf {\bibinfo {volume} {82}},\ \bibinfo {pages} {52} (\bibinfo
  {year} {2003})}\BibitemShut {NoStop}%
\bibitem [{\citenamefont {Bailey}\ \emph {et~al.}(2008)\citenamefont {Bailey},
  \citenamefont {Veal}, \citenamefont {King}, \citenamefont {McConville},
  \citenamefont {Pereiro}, \citenamefont {Grandal}, \citenamefont
  {S{\'a}nchez-Garc{\'\i}a}, \citenamefont {Mu{\~n}oz},\ and\ \citenamefont
  {Calleja}}]{bailey_joap08}%
  \BibitemOpen
  \bibfield  {author} {\bibinfo {author} {\bibfnamefont {L.~R.}\ \bibnamefont
  {Bailey}}, \bibinfo {author} {\bibfnamefont {T.~D.}\ \bibnamefont {Veal}},
  \bibinfo {author} {\bibfnamefont {P.}~\bibnamefont {King}}, \bibinfo {author}
  {\bibfnamefont {C.~F.}\ \bibnamefont {McConville}}, \bibinfo {author}
  {\bibfnamefont {J.}~\bibnamefont {Pereiro}}, \bibinfo {author} {\bibfnamefont
  {J.}~\bibnamefont {Grandal}}, \bibinfo {author} {\bibfnamefont
  {M.}~\bibnamefont {S{\'a}nchez-Garc{\'\i}a}}, \bibinfo {author}
  {\bibfnamefont {E.}~\bibnamefont {Mu{\~n}oz}}, \ and\ \bibinfo {author}
  {\bibfnamefont {E.}~\bibnamefont {Calleja}},\ }\href@noop {} {\bibfield
  {journal} {\bibinfo  {journal} {J. Appl. Phys.}\ }\textbf {\bibinfo {volume}
  {104}},\ \bibinfo {pages} {113716} (\bibinfo {year} {2008})}\BibitemShut
  {NoStop}%
\bibitem [{\citenamefont {King}\ \emph {et~al.}(2008)\citenamefont {King},
  \citenamefont {Veal}, \citenamefont {Lu}, \citenamefont {Jefferson},
  \citenamefont {Hatfield}, \citenamefont {Schaff},\ and\ \citenamefont
  {McConville}}]{king_pssb08}%
  \BibitemOpen
  \bibfield  {author} {\bibinfo {author} {\bibfnamefont {P.}~\bibnamefont
  {King}}, \bibinfo {author} {\bibfnamefont {T.}~\bibnamefont {Veal}}, \bibinfo
  {author} {\bibfnamefont {H.}~\bibnamefont {Lu}}, \bibinfo {author}
  {\bibfnamefont {P.~H.}\ \bibnamefont {Jefferson}}, \bibinfo {author}
  {\bibfnamefont {S.}~\bibnamefont {Hatfield}}, \bibinfo {author}
  {\bibfnamefont {W.~J.}\ \bibnamefont {Schaff}}, \ and\ \bibinfo {author}
  {\bibfnamefont {C.}~\bibnamefont {McConville}},\ }\href@noop {} {\bibfield
  {journal} {\bibinfo  {journal} {Phys. Status Solidi B}\ }\textbf {\bibinfo
  {volume} {245}},\ \bibinfo {pages} {881} (\bibinfo {year}
  {2008})}\BibitemShut {NoStop}%
\bibitem [{\citenamefont {Fu}\ and\ \citenamefont {Chen}(2004)}]{Fu_1}%
  \BibitemOpen
  \bibfield  {author} {\bibinfo {author} {\bibfnamefont {S.~P.}\ \bibnamefont
  {Fu}}\ and\ \bibinfo {author} {\bibfnamefont {Y.~F.}\ \bibnamefont {Chen}},\
  }\href {\doibase http://dx.doi.org/10.1063/1.1787615} {\bibfield  {journal}
  {\bibinfo  {journal} {Appl. Phys. Lett.}\ }\textbf {\bibinfo {volume} {85}},\
  \bibinfo {pages} {1523} (\bibinfo {year} {2004})}\BibitemShut {NoStop}%
\end{thebibliography}%
\end{document}